\documentclass[12pt,a4]{article}   
\usepackage{amsmath,amssymb}  
\def \be {\begin{equation}}   
\def \ee {\end{equation}}   
\def \bea  {\begin{eqnarray}}   
\def \eea {\end{eqnarray}}   
   
\title{ Complex numbers and symmetries in quantum mechanics,
and a nonlinear superposition principle for Wigner functions}   
\author{ A.J. Bracken \\ Department of Mathematics\\and Centre for Mathematical Physics \\  
University of Queensland\\Brisbane\\Queensland 4072\\Australia\\  
ajb@maths.uq.edu.au}   
   
\begin{document}   
   
\maketitle   
\begin{abstract}   
Complex numbers appear in the Hilbert space formulation of quantum mechanics, but not in the
formulation in phase space.  Quantum symmetries are described by complex, unitary or antiunitary
operators defining ray representations in Hilbert space, whereas 
in phase space they are described by real,
true representations.  Equivalence of the formulations requires that the former representations
can be obtained from
the latter and {\em vice versa}.  Examples are given.  Equivalence of the two formulations also
requires that complex superpositions
of state vectors can be described in the phase space formulation, and it is
shown that this leads to a nonlinear superposition principle for orthogonal, pure-state Wigner
functions.  It is concluded that the use of complex numbers in quantum mechanics can be regarded as
a computational
device to simplify calculations, 
as in all other applications of mathematics to physical phenomena.
\end{abstract}   
   
\def \ni {\noindent}  
\def \vs  {\vskip5mm}

\section{Introduction}  
  
\vs\ni  
One of the remarkable features of quantum mechanics as usually formulated is the fundamental role  
played there by the complex numbers.  
This appears to distinguish quantum mechanics from other mathematical  
models of natural phenomena. Theoreticians use complex numbers routinely in  
most  
applications of mathematics, starting with  
the solution of simple quadratic equations.  
However, in all other models, the complex numbers are introduced only as a computational tool.  
Not only are the `observables' of these models real but, invariably,  so also are   
their defining  equations.

Feynman  and Dirac were prominent among those who have stressed the  
crucial role  of complex probability amplitudes in quantum mechanics.

To quote  
Feynman:  
{\it It has been found that all processes so far observed can be understood in terms of the following  
prescription: To every process there corresponds an amplitude (a complex number); with proper  
normalization the probability of the process is equal to the absolute square of the amplitude}  
\cite{feynman}.

And Dirac:  
{\it So if one asks what is the main feature of quantum mechanics, I feel inclined now to say that it is  
... the existence of probability amplitudes which underly all atomic processes.   Now a probability  
amplitude is related to experiment but only partially.  The square of its modulus is something that  
we can observe.  That is the probability which the experimental people get.  But besides that there  
is a phase ... (which) is all important because it is the source of all interference phenomena but  
its physical meaning is obscure}  
\cite{dirac}.

The essential place of complex numbers in quantum mechanics  
as usually formulated  
is evident in  Schr\"odinger's time-dependent wave equation  
\begin{equation}  
i\hbar\frac{\partial \psi(t)}{\partial t}={\hat H}\psi(t)\,,  
\label{schrodinger}  
\end{equation}  
where $\psi(t)$ is the state vector of a quantum system and ${\hat H}$ is its Hamiltonian operator.  
For spinless, nonrelativistic systems, ${\hat H}$ is a real operator in   
the usual coordinate realization of Hilbert space, and the   
radical departure   
that the complex equation (\ref{schrodinger}) then represents,   
from the reality of the governing equations of all earlier models of  
natural phenomena,   
was a departure   
that Schr\"odinger made uneasily  
\cite{moore}.

There is a formulation of quantum mechanics \cite{groenewold}, 
\cite{moyal}, \cite{dubin} -- I will refer to it as the  
phase-space formulation -- that has  become popular as   
an arena in which to investigate  such matters as   
quantum chaos \cite{berry}, quantum tomography \cite{leonhardt}, \cite{doebner},  
the relationship between quantum and  
classical mechanics \cite{osborn}, \cite{bracken}, and the nature of quantization   
\cite{bayen}, \cite{fronsdal}.    
What I would like to draw attention to here is   
that the phase-space formulation is purely real -- there are no complex  
numbers to be seen in the defining equations. Despite this, it has been   
stated \cite{bayen}, \cite{zachos} that the phase-space  
formulation is   
equivalent to the more familiar formulation in terms of hermitian operators acting on  
a  complex Hilbert space.    
In view of the remarks above, one may well ask how this can  be so.

In what follows, I shall discuss two aspects of this surprising state of affairs.    
The first concerns representations  of symmetry groups and algebras.    
In the Hilbert space formulation,  elements of a symmetry group or more generally, of any group of  
automorphisms of the set of quantum states,   
are represented by unitary or antiunitary operators  
that define ray representations   
of the group as a whole, in accordance with  
Wigner's celebrated theorem \cite{wigner,bargmann,cassinelliB}.    
In contrast, in the phase-space formulation,  elements of a symmetry  
group are represented by real  
unitary operators that define a true representation.  
If the phase-space formulation is equivalent to the one 
in Hilbert space, then the complex ray representations must  
be recoverable from the real true ones, and {\em vice versa}. 
That this is 
indeed the case \cite{cassinelli} is remarkable from the point of view of the  
mathematics and the physics involved.  
  
The second aspect concerns the fact that in the phase-space 
formulation, quantum states are described by real
Wigner functions, whereas in Hilbert  space, one has 
state vectors that can be superposed, and that have  
complex phases.      
We shall see that there is a corresponding  
nonlinear superposition principle for pure-state Wigner functions,  that incorporates  
the relative  
phase of state vectors that are supersposed.

\section{Group representations}  
  
Two examples for systems with one linear degree of freedom   
illustrate  the strikingly different treatment  
of  
group representations in the Hilbert space  
and phase space formulations \cite{cassinelli}.

Consider firstly  
the Heisenberg-Weyl group, which in the abstract  may be considered as the 2-parameter abelian  
Lie  
group of transformations $g(a,b)$, $a,\,b\in \mathbb R$,  acting on   
real, square-integrable functions $F(q,p)$ of two real variables as  
\begin{equation}  
g(a,b)\,:\quad F(q,p)\longrightarrow F(q+a,p-b) \,,  
\label{abelianweylgroup}  
\end{equation}  
We can write, on suitably smooth $F$,   
\begin{eqnarray}  
g(a,b) = \exp[a x_1 +b x_2]\,,  
\quad g(a,b)\,g(c,d)=g(a+c, b+d)\,.  
\nonumber\\  
\nonumber\\  
x_1=\partial/\partial q\,,\quad x_2=-\partial/\partial p\,, \quad 
\left[x_1\,,\,x_2 \right] =0\,.  \qquad\qquad
\label{abelianweylalgebra}  
\end{eqnarray}  
and we note the appearance of the 2-dimensional real Lie algebra spanned by $x_1$ and $x_2$.    
  
The familiar unitary ray representation  
of this group on Hilbert space has   
\begin{eqnarray}  
g(a,b)\longrightarrow U(a,b)\,,\qquad\qquad\qquad  
\nonumber\\  
\nonumber\\  
U(a,b)U(c,d)\,=\,e^{i(ad-bc)/2\hbar}\,U(a+c,b+d)\,.    
\label{quantumweyl}  
\end{eqnarray}  
With   
\begin{equation}  
U(a,b)=  
\exp[i(a{\hat P} + b{\hat Q})/\hbar]\,,  
\label{Uform}  
\end{equation}  
we have  
\begin{equation}  
\left[ {\hat Q}\,,\, {\hat P}\right]= i\hbar {\hat I}\,  
\label{heisenbergalgebra}  
\end{equation}  
where ${\hat I}$ is the unit operator,   
so that at the level of the Lie algebra the ray representation of the group is associated with  
a central extension, characterised by Planck's  
constant $\hbar$.  In either the coordinate realization or the momentum realization of Hilbert space,  
one of the  
operators ${\hat Q}$, ${\hat P}$ is real, the other pure imaginary, and again we see the essential role  
played by the complex numbers.    
  
Because this ray representation and central extension lie at the very heart of the Hilbert space  
formulation, it is all the more surprising to find that in the phase space formulation, the  
Heisenberg-Weyl group simply has its defining  representation  
(\ref{abelianweylgroup},  
\ref{abelianweylalgebra}).   
This representation is real and true; not only the complex numbers, but also Planck's constant, are  
nowhere to be seen.    
  
The second example is provided by the time reversal group of order 2, with elements $g$, $e$ acting   
on functions $F(q,\,p)$ as  
\begin{eqnarray}  
g\,:F(q,\,p) \longrightarrow F(q,\,-p)\,,\quad e\,:F(q,\,p)\longrightarrow F(q,\,p)\,,    
\quad (\,\Rightarrow g^2=e\,)\,.  
\label{timereversal}  
\end{eqnarray}  
  
In the Hilbert space formulation, $e$ is represented by  
the unit operator ${\hat U}_e= {\hat I}$, and $g$ by an   
antiunitary operator ${\hat A}_g$.  In the coordinate realization, for example, we may have   
\begin{eqnarray}  
{\hat A}_g\,: \psi(x)\longrightarrow \psi^{*}(x)\,,&&\quad {\hat U}_e\,: \psi(x)\longrightarrow \psi(x)\,,  
\quad (\,\Rightarrow A_g^2=U_e\,)\,.  
\label{timereversalrep}  
\end{eqnarray}  
Once again, the representation in the phase space formulation is simply  the defining  
representation (\ref{timereversal}), and in sharp contrast to the Hilbert space representation, is real  
and   
unitary.   
  
It is not hard to find the source of these differences.    
The real structure that is inherent in the phase space formulation   
is in fact already present in the Hilbert  space  
formulation.  To see this, we recall \cite{dubin}, \cite{cassinelli} that the   
mapping from the Hilbert space formulation   
to the phase space formulation is the   
Weyl-Wigner transform ${\cal W}$, whose inverse ${\cal W}^{-1}$ is Weyl's quantization map.     
The transform ${\cal W}$ does not act on state vectors in Hilbert space at all.  Rather it   
maps the real Hilbert space ${\cal T}$  
of hermitian Hilbert-Schmidt operators ${\hat A}$ that act   
on the space of state vectors,   
onto the real Hilbert space ${\cal K}$ of square integrable functions $A$   
on phase space.  Conversely,   
${\cal W}^{-1}$ carries such  
functions $A$ back into hermitian operators ${\hat A}$ (that is, it quantizes them).   
If $A_K(x,y)$ is the kernel  
of ${\hat A}$, regarded as an integral operator   
in the coordinate realization  
of Hilbert space, then the action of ${\cal W}$ and ${\cal W}^{-1}$ is defined by  
\begin{eqnarray}  
{\cal W}({\hat A})(q,\,p)=\int A_K(q-y/2, q+y/2)\,e^{ipy/\hbar}\,dy\,,  
\nonumber\\  
\nonumber\\  
{\cal W}^{-1}(A)_K(x,y)=\frac{1}{2\pi\hbar}\int A([x+y]/2,\, p)\,e^{ip(x-y)/\hbar}\,dp\,.  
\label{WWinvdef}  
\end{eqnarray}  
With a suitably generalized interpretation of the integrals, these formulas also   
define the action of ${\cal W}$ on hermitian   
operators that lie   
outside ${\cal T}$, such as ${\hat Q}$, ${\hat P}$, and the action of ${\cal W}^{-1}$  
on real-valued phase space functions that lie outside ${\cal K}$,  
such as $q$ and $p$.    
   
It can be seen from this that the question of the equivalence of the Hilbert space and phase space  
formulations of quantum mechanics  
is really a question as to whether or not, in the usual Hilbert space formulation, one needs  
state vectors to truly describe states, or whether it is   
sufficient to work only with  density operators and the  
real algebra of hermitian operators.    
  
The density operator ${\hat  
\rho}$ for the state of  
the system involved, lies in ${\cal T}$ and is mapped by ${\cal W}$ into   
$(2\pi\hbar\times)$ Wigner's pseudo probability density    
function $W_{\rho}\in{\cal K}$ on phase space.  
But there is no mapping of the complex state  
vector, which has no image appearing in the phase space formulation.     
  
A ray representation ${\hat \pi}$  
of a group $G$, defined by unitary and antiunitary operators acting on state vectors $\psi$, also  
defines a  
real, true representation $\Pi_{\cal T}$ of $G$ on   
${\cal T}$,     
\begin{eqnarray}  
g\in G\,: \, \psi\longrightarrow {\hat \pi}(g)\,\psi\,,  
\nonumber\\  
\nonumber\\  
{\hat A}\longrightarrow \Pi_{\cal T}(g)({\hat A})= {\hat \pi}(g)\,{\hat A}\,{\hat \pi}(g)^{\dagger}\,.  
\label{repA}  
\end{eqnarray}  
It is this real, true representation $\Pi_{\cal T}$   
that is mapped by the Weyl-Wigner transform onto a real, true  
representation $\Pi_{\cal K}$ on square-integrable functions on phase space.    
  
From (\ref{repA}) it can be seen that $\Pi_{\cal K}$ is isomorphic to the tensor product of the ray  
representation ${\hat \pi}$ with its contragredient ${\hat \pi}^C$; elsewhere \cite{cassinelli}  
we have called $\Pi_{\cal  
K}$ the Weyl-Wigner product of ${\hat \pi}$ and ${\hat \pi}^C$.    
  
This explains  
why groups have real, true representations in the phase-space formulation.    
But if that formulation is  
indeed equivalent to the Hilbert space formulation, it   
must be possible to recover the underlying ray  
representations from the phase-space ones.    
In effect, we must be able to `factorize' $\Pi_{\cal K}$, or  
$\Pi_{\cal T}$, and determine ${\hat \pi}$.  That this is   
possible is surprising, especially when it is  
recalled that at the level of the Lie algebra we have to   
determine a centrally extended representation,  
with associated parameter(s),   
from a true one.  Because the representation $\Pi_{\cal K}$ can   
typically be regarded as the defining  
representation of the group of interest, as in the examples of the Heisenberg-Weyl group and
time reversal group above, it can be said that in finding ${\hat \pi}$   
we are expressing Wigner's  
Theorem in a constructive way for that group. 

It has to   
be emphasized that not every real, true  
representation on ${\cal K}$ of a given group can be factorized in this way.    
Necessary and sufficient  
conditions for this to be possible have been determined  
elsewhere \cite{cassinelli}.    
We shall expand here on one aspect only: the   
construction of ${\hat \pi}$ in the case of a 1-parameter  Lie group,   
whose representation $\Pi_{\cal K}$  
is generated by a real, skew-adjoint operator on ${\cal K}$,   
which we may think of as an integral operator  
${\hat \alpha}$ with real kernel $\alpha_{\cal K}$,  
\begin{equation}  
\big({\hat \alpha} F\big)(q,\,p)=  
\int \alpha_{\cal K} (q,\,p,\,q',\,p')\,F(q',\,p')\,dq '\,dp'\,.  
\label{alphaaction}  
\end{equation}  
  
In this case, the necessary and sufficient condition on ${\hat \alpha}$ for   
factorizability is \cite{cassinelli}  
  
\begin{eqnarray}  
\int \sin[(qp_1+pq_1)/\hbar]\,R(q_1,\,p_1,\,q',\,p')\,dq_1\,dp_1\qquad\qquad\qquad\qquad\qquad  
\nonumber\\  
\nonumber\\  
=\int\sin[(q_1p+q_1p'+p_1q+p_1q')/2\hbar]\,R(q_1,\,p_1,\,[q+q']/2,\,[p+p']/2)\,dq_1  \,dp_1
\nonumber\\  
\nonumber\\  
-\int\sin[(q_1 p-q_1p'+p_1q'-p_1q)/2\hbar]\,R(q_1,\,p_1,\,[q'-q]/2,\,[p'-p]/2)\,dq_1\,dp_1\,,  
\label{necesssuffcondition}  
\end{eqnarray}  
where  
\begin{equation}  
R(q,\,p,\,q',\,p')=\alpha_{\cal K}([q'-q]/2,\,[p'+p]/2,\,[q'+q]/2,\,[p'-p]/2)\,.  
\label{Rform}  
\end{equation}  
When this condition holds, we can obtain the (complex) hermitian generator   
${\hat A}$
of the representation ${\hat \pi}$ on  
Hilbert space by setting  
\begin{equation}  
A(q,\,p)=  
c+\int \sin[ (qp'+pq')/\hbar]  
\,R(q',\,p',\,q,\,p)\, dq'\,dp'\,,  
\label{Adef}  
\end{equation}  
where $c$ is an arbitrary constant, and then  
\begin{equation}  
{\hat A}={\cal W}^{-1}(A)\,.  
\label{Ahatdef}  
\end{equation}  
Note that ${\hat A}$ is defined only up to the addition of a multiple of  
${\hat I}$ by $c$, leading to an arbitrary overall phase in the   
representation of the 1-parameter group on Hilbert space.    
  
It is surprising that complex ray   
representations can be obtained from true representations   
by this factorization process, which involves the `quantization' step   
(\ref{Ahatdef}).  At the level of the   
corresponding Lie algebra,  
we may obtain a representation of a central extension   
of that algebra in this way.   
Clearly any extension parameters must be built   
into the transforms (\ref{necesssuffcondition})  
and ${\cal W}^{-1}$ (namely $\hbar$), or else appear already in the   
true representation on phase space.   
Both possibilities arise in practice \cite{cassinelli}.   
  
As an example, consider again the Heisenberg-Weyl group for one degree of freedom,  
with representation (\ref{abelianweylgroup}, \ref{abelianweylalgebra}) on phase space.  
In this case, as seen from (\ref{abelianweylalgebra}),  $\alpha_1=x_1=\partial/\partial q$,   
$\alpha_2=x_2=-\partial/\partial p$, leading from (\ref{alphaaction}) to  
  
\begin{equation}  
\alpha_{1K}(q,\,p,\,q',\,p')=\delta '(q-q')\delta(p-p')\,,\quad  
\alpha_{2K}(q,\,p,\,q',\,p')=-\delta(q-q')\delta '(p-p')\,,  
\label{alphaforms}  
\end{equation}  
and hence from (\ref{Adef}) and (\ref{Ahatdef}) to  
\begin{eqnarray}  
A_1(q,\,p)=p+c_1\,,\quad A_2(q,\,p)=q+c_2\,,  
\nonumber\\  
\nonumber\\  
{\hat A}_1={\hat P}+c_1{\hat I}\,,\quad  
{\hat A}_2={\hat Q}+c_2{\hat I}\,. \qquad 
\label{AAhatforms}  
\end{eqnarray}   
The constant terms in ${\hat A}_1$, ${\hat A}_2$
can be removed by a unitary transformation generated by ${\hat Q}$   
followed by a unitary transformation generated by ${\hat P}$, leaving  
${\hat P}$ and ${\hat Q}$  satisfying (\ref{heisenbergalgebra}).

\section{Superposition of Wigner functions}  
According to the first of (\ref{WWinvdef}), the Wigner function $W_{\rho}$ corresponding to   
$([2\pi\hbar]^{-1} \times)$   
the pure-state   
density operator ${\hat \rho}$ with kernel $\rho_K(x,y)=\varphi(x)\varphi^{*}(y)$ is    
\begin{equation}  
W_{\rho}(q,\,p)=\frac{1}{2\pi\hbar}\int \varphi(q-y/2)\varphi^{*}  
(q+y/2)\,e^{ipy/\hbar}  
\,dy\,.  
\label{wigfndef}  
\end{equation}  
Here $\varphi (x)$ is the representative  of a normalized pure state vector  
corresponding to ${\hat \rho}$,    
in the coordinate realization of Hilbert space.  It  is clear that $\varphi(x)$ can be   
recovered from $W_{\rho}(q,\,p)$ by applying the second of 
(\ref{WWinvdef}).  We have   \cite{tatarskii}
\begin{equation}  
\varphi(x)\varphi^{*}(x_0)=\int W_{\rho}([x+x_0]/2,\,p)\,e^{ip(x-x_0)/\hbar}\,dp\,,  
\label{phiformula}  
\end{equation}  
which determines a fixed complex   
multiple of $\varphi(x)$ so long as we choose any particular $x_0$ such that   
$\varphi(x_0)\neq 0$ and hence such that the left-hand-side of (\ref{phiformula})   
is not identically equal to zero. What we wish to show here is that a complex superposition of   
two state vectors,  
including their relative phase, can also be described in terms of the   
two corresponding real Wigner functions.      
  
Suppose then that we are given two normalized, orthogonal pure-state wave functions,  
$\varphi_1(x)$, $\varphi_2(x)$, from which we have constructed  
the corresponding density operators   
${\hat \rho}_1$, ${\hat \rho}_2$, and the corresponding  
Wigner functions $W_1(q,\,p)$, $W_2(q,\,p)$.  
We consider the normalized superposition  
\begin{equation}  
\varphi (x) =a_1 \,\varphi_1(x) + a_2 \,\varphi_2(x)\,,\quad |a_1|^2 + |a_2|^2=1\,.  
\label{superpositionA}  
\end{equation}  
Choosing particular values $x_1$, $x_2$ such that   
$\varphi_1(x_1)\neq 0$, $\varphi_2(x_2)\neq 0$,  
we rewrite (\ref{superpositionA}) as  
\begin{equation}  
\varphi(x)=c_1 \,\varphi_1(x)\varphi_1^{*}(x_1)+  
c_2\,\varphi_2(x)\varphi_2^{*}(x_2)\,,\quad  
|c_1|^2|\varphi_1(x_1)|^2  
+  
|c_2|^2|\varphi_2(x_2)|^2 =1\,.  
\label{superpositionB}  
\end{equation}  
The kernel of the density operator corresponding to the pure state $\varphi (x)$ is  
then   
\begin{eqnarray}  
\rho_K(x,y)=\varphi(x)\varphi^{*}(y)=|a_1|^2\varphi_1(x)\varphi_1^{*}(y)  
+  
|a_2|^2\varphi_2(x)\varphi_2^{*}(y)\qquad\qquad\qquad\qquad  
\nonumber\\  
\nonumber\\  
+c_1c_2^{*}\varphi_1(x)\varphi_1^{*}(x_1)\varphi_2^{*}(y)\varphi_2(x_2)  
+c_2c_1^{*}\varphi_2(x)\varphi_2^{*}(x_2)\varphi_1^{*}(y)\varphi_1(x_1)\,,  
\nonumber\\  
\nonumber\\  
=|a_1|^2\,\rho_{1K}(x,y)+  
|a_2|^2\,\rho_{2K}(x,y)\qquad\qquad\qquad\qquad\qquad\qquad\qquad  
\nonumber\\  
\nonumber\\  
+  
c_1c_2^{*}\rho_{1K}(x,x_1)\rho_{2K}^{*}(y,x_2)+  
c_2c_1^{*}\rho_{2K}(x,x_2)\rho_{1K}^{*}(y,x_1)\,.  
\label{superpositionC}  
\end{eqnarray}  
Applying the first of (\ref{WWinvdef}) to $\rho_K/2\pi\hbar$, 
we obtain

\begin{eqnarray}  
W_{\rho}(q,\,p)=|a_1|^2\,W_1(q,\,p)+|a_2|^2\,W_2(q,\,p)  
\nonumber\\  
\nonumber\\  
+\frac{c_1c_2^{*}}{2\pi\hbar} 
\int\rho_{1K}(q-y/2),x_1)\rho_{2K}^{*}(q+y/2,x_2)\,e^{ipy/\hbar}\,dy  
\nonumber\\  
\nonumber\\  
+\frac{c_2c_1^{*}}{2\pi\hbar} 
\int\rho_{2K}(q-y/2),x_2)\rho_{1K}^{*}(q+y/2,x_1)\,e^{ipy/\hbar}\,dy\,.  
\label{superpositionD}  
\end{eqnarray}  
Next rewriting $\rho_{1K}$ and $\rho_{2K}$ using the second of (\ref{WWinvdef}),  
we see that the third term on the RHS of (\ref{superpositionD})  
can be written as  
\begin{eqnarray}  
\frac{c_1 c_2^{*}}{2\pi\hbar} 
\int\!\!\!\int\!\!\!\int  
W_1([q-y/2+x_1]/2, \,p_1)W_2([q+y/2+x_2]/2,\,p_2) 
\nonumber\\ 
\nonumber\\ 
\times\,e^{ip_1[q-y/2-x_1]/\hbar}\,e^{-ip_2[q+y/2-x_2]/\hbar}\,e^{ipy/\hbar}\,dp_1\,dp_2\,dy\,,  
\label{superpositionE}  
\end{eqnarray}  
where we have noted the reality of $W_2$.  Because the   
fourth term on the RHS of  (\ref{superpositionD}) is the complex   
conjugate of the third, we see that   
the sum of the third and fourth terms can be written as 
\begin{eqnarray} 
\frac{|c_1||c_2|}{\pi\hbar}\int\!\!\!\int\!\!\!\int  
W_1([q-y/2+x_1]/2, \,p_1)W_2([q+y/2+x_2]/2,\,p_2) 
\nonumber\\ 
\nonumber\\ 
\times\,\cos[\epsilon +(2p -p_1-p_2)y/2+(p_1-p_2)q -p_1x_1+p_2x_2]\,dp_1\,dp_2\,dy\,, 
\label{superpositionF} 
\end{eqnarray} 
where we have written 
\begin{equation} 
c_1c_2^{*}=|c_1||c_2|e^{i\epsilon}\,,\quad 0\leq\epsilon < 2\pi\,. 
\label{epsilondef} 
\end{equation} 
Next we note that 
\begin{eqnarray} 
|a_1|^2=|c_1|^2 |\varphi_1(x_1)|^2=|c_1|^2\,\int W(x_1,\,p)\,dp\,, 
\label{coeffrelnA} 
\end{eqnarray} 
with a similar relation for $|a_2|^2$.  Then we can write (\ref{superpositionD}) as 
\begin{eqnarray} 
W(q,\,p)= 
|c_1|^2\big\{\int\,W(x_1,\,p')\,dp'\big\}\,W_1(q,\,p) 
+ 
|c_2|^2\big\{\int\,W(x_2,\,p')\,dp'\big\}\,W_2(q,\,p) 
\nonumber\\ 
\nonumber\\ 
+ 
\frac{|c_1||c_2|}{\pi\hbar}\int\!\!\!\int\!\!\!\int  
W_1([q-y/2+x_1]/2, \,p_1)W_2([q+y/2+x_2]/2,\,p_2) 
\nonumber\\ 
\nonumber\\ 
\times\,\cos[\epsilon +(2p -p_1-p_2)y/2+(p_1-p_2)q -p_1x_1+p_2x_2]\,dp_1\,dp_2\,dy\,. 
\label{superpositionG} 
\end{eqnarray} 
 
This is a nonlinear superposition rule for Wigner functions, expressing a new Wigner function $W$  
in terms of two 
given ones, $W_1$ and $W_2$, two nonnegative 
coefficients $|c_1|$, $|c_2|$, and a phase angle $\epsilon$ between 
$0$ and $2\pi$ as in (\ref{epsilondef}).    
The two coordinates $x_1$, $x_2$ in (\ref{superpositionF}) are arbitrary, except that 
we must have 
\begin{eqnarray} 
\int\,W_1(x_1,\,p)\,dp\neq 0,\quad  
\int\,W_2(x_2,\,p)\,dp\neq 0,\, 
\label{coordcondition} 
\end{eqnarray} 
for consistency with our assumption above that $\varphi(x_1)\neq 0$ and $\varphi(x_2)\neq 0$.  Since 
we also assumed that $\varphi_1$ and $\varphi_2$ are orthogonal, we must also require that $W_1$ and 
$W_2$ are orthogonal, that is 
\begin{eqnarray} 
\int\!\!\!\int\,W_1(q,\,p)\,W_2(q,\,p)\,dq\,dp=\frac{1}{2\pi\hbar} 
\big|\int\varphi_1^{*}(x)\varphi_2(x)\,dx\big |^2=0\,, 
\label{Worthog} 
\end{eqnarray} 
and since we assumed that $W_1$ and $W_2$ correspond to pure states,  
they must also satisfy the pure-state conditions \cite{tatarskii}
\begin{equation} 
\int W_1(q,\,p)^2\,dq\,dp=\frac{1}{2\pi\hbar}\,,\quad 
\int W_2(q,\,p)^2\,dq\,dp=\frac{1}{2\pi\hbar}\,, 
\label{pure_wigner} 
\end{equation} 
as well as the normalization conditions 
\begin{equation} 
\int W_1(q,\,p)\,dq\,dp=1\,,\quad 
\int W_2(q,\,p)\,dq\,dp=1\,. 
\label{wigner_normalize} 
\end{equation} 
 
It can now be seen that (\ref{superpositionA}) and (\ref{superpositionG}) are equivalent.   
If we are given two orthogonal, normalized wave functions 
$\varphi_1$ and $\varphi_2$ and form the superposition (\ref{superpositionA}),  
then from $\varphi$ we can construct  
$W_{\rho}$ as in (\ref{superpositionG}) following the steps outlined above.   
Conversely, if we start with $W_1$ and $W_2$, we  
can find an $x_1$ and $x_2$ such that 
(\ref{coordcondition}) hold, and form the nonlinear superposition (\ref{superpositionG}) for some  
choice of $|c_1|$, $|c_2|$ and $\epsilon$ such that 
the second of (\ref{epsilondef}) is satisfied.  Note that fixing $|c_1|$, $|c_2|$  
and $\epsilon$ fixes complex $c_1$ and $c_2$, 
up to an overall phase, as for example in (\ref{coeffdefB}) below.  
We can also construct $\varphi_1(x)\varphi_1^{*}(x_1)$  
and $\varphi_2(x)\varphi_2^{*}(x_2)$ from  
$W_1$ and $W_2$, as in (\ref{phiformula}), and then recover $\varphi$  
as in (\ref{superpositionA}), up to an overall phase.   
 
The result  can also be expressed as follows.   
Given two orthogonal pure-state Wigner functions $W_1$ and $W_2$ 
satisfying (\ref{Worthog}), (\ref{pure_wigner}) and (\ref{wigner_normalize}), we choose any  
constants $x_1$ and $x_2$ such that (\ref{coordcondition}) hold, and  
any constants $A$, $B$ and $\epsilon$ satisfying 
\begin{equation} 
A\geq 0\,,\quad B\geq 0\,,\quad 0\leq \epsilon < 2\pi\,, 
\label{ABepsilon} 
\end{equation} 
and we form 
\begin{eqnarray} 
T(q,\,p)= 
A^2\big\{\int\,W(x_1,\,p')\,dp'\big\}\,W_1(q,\,p) 
+ 
B^2\big\{\int\,W(x_2,\,p')\,dp'\big\}\,W_2(q,\,p) 
\nonumber\\ 
\nonumber\\ 
+ 
\frac{AB}{\pi\hbar}\int\!\!\!\int\!\!\!\int  
W_1([q-y/2+x_1]/2, \,p_1)W_2([q+y/2+x_2]/2,\,p_2) 
\nonumber\\ 
\nonumber\\ 
\times\,\cos[\epsilon +(2p -p_1-p_2)y/2+(p_1-p_2)q -p_1x_1+p_2x_2]\,dp_1\,dp_2\,dy\,. 
\label{superpositionH} 
\end{eqnarray} 
Then 
\begin{equation} 
W(q,\,p)=T(q,\,p)/\int T(q',\,p')\,dq'\,dp' 
\label{wignerform} 
\end{equation} 
is a normalized pure state Wigner function.  It corresponds to  
the wave function superposition (\ref{superpositionB}), 
where $\varphi_1(x)\varphi_1^{*}(x_1)$ and $\varphi_2(x)\varphi_2^{*}(x_2)$  
are constructed from $W_1$ and $W_2$ as 
in (\ref{phiformula}), 
and  
\begin{eqnarray} 
|c_1|=A/\big\{\int T(q,\,p)\,dq\,dp\big\}^{1/2}\,,\quad 
|c_2|=B/\big\{\int T(q,\,p)\,dq\,dp\big\}^{1/2}\,,\ 
\nonumber\\ 
\nonumber\\ 
c_1=|c_1|\,,\quad c_2=|c_2|\,e^{-i\epsilon}\,. 
\qquad\qquad\qquad\qquad
\label{coeffdefB}
\end{eqnarray} 
 
\section{Conclusions} 
Quantum symmetries and complex superpositions of pure, orthogonal quantum states,  
can both be described in the phase space formulation of quantum mechanics, 
without the use of complex numbers.  The description of symmetries is much simpler in phase space, 
but the description of superpositions is much more 
complicated.  The description of non-orthogonal superpositions 
is even more complicated, and has not been attempted here. 
 
Note that the relative phase of the superposed wave functions, whose 
{\it physical meaning is obscure} \,\cite{dirac}, is essentially the $\epsilon$ appearing in 
the nonlinear superposition formulas (\ref{superpositionG}) and (\ref{superpositionH}).  
 
We conclude that the phase space formulation does indeed appear capable of  
reproducing all aspects of quantum mechanics. 
In the case of the superposition of quantum states however, this is only be achieved at the cost of 
much greater 
complication.   
 
If we wish to think of the phase space formulation as the more fundamental, arising directly from  
a deformation of classical mechanics in phase space \cite{bayen}, we can think of 
the formulation of quantum mechanics 
in Hilbert space, and the associated introduction of complex numbers, as a computational device to  
make calculations easier.  
From this point of view, the apearance of complex numbers in quantum mechanics
is on a similar footing to their appearance in other applications 
of mathematics to natural phenomena.


\begin{thebibliography}{99}  
  

  
\bibitem{feynman} Feynman, R.P., {\em The Theory of Fundamental Processes}   
(Benjamin/Cummings,  
Reading, Mass., 1962).  
 
\bibitem{dirac} Dirac, P.A.M., {\em Fields and Quanta} {\bf 3} (1972), 139--164.       
  
\bibitem{moore} Moore, W.J., {\em Schr\"odinger: Life and Thought}   
(Cambridge University Press, Cambridge,  
England,  
1989).  
   
\bibitem{groenewold} Groenewold, H., {\em Physica} {\bf 12} (1946), 405--460.

\bibitem{moyal} Moyal, J.E., {\em Proc. Camb. Phil. Soc.} {\bf 45} (1949), 99--124. 
 
\bibitem{dubin} Dubin, D.A., Hennings, M.A. and Smith, T.B., {\em Mathematical Aspects of 
Weyl Quantization and Phase} (World Scientific, Singapore, 2000). 
 
\bibitem{berry} Berry, M.V., {\em Phil. Trans. R. Soc. A} {\bf 287} (1977), 237--271. 
 
\bibitem{leonhardt} Leonhardt, U., {\em Measuring the Quantum State of Light}  
(Cambridge University Press, Cambridge, England, 1997). 
 
\bibitem{doebner} Bracken, A.J., Doebner, H.-D. and Wood, J.G., {\em Phys. Rev. Lett.} 
 {\bf 83} 
(1999), 3758--3761.  
 
 
\bibitem{osborn}Osborn, T.A. and Molzahn, F.H., {\em Annals Phys.} {\bf 241} (1995), 79--127. 
 
\bibitem{bracken} Bracken, A.J., {\em J. Phys. A} {\bf 36} (2003), L329--L335.
 
\bibitem{bayen} Bayen, F., Flato, M., Fronsdal, C., Lichnerowicz, A. and Sterheimer, D., 
{\em Annals Phys.} {\bf 111} (1978), 61--110; 111--151. 
 
\bibitem{fronsdal} Fronsdal, C., {Rep. Math. Phys.} {\bf 15} (1978), 111--145. 
 
 
\bibitem{zachos} Zachos, C. K., {\em Int. J. Mod. Phys. A} {\bf 17} (2002), 297--316. 

 
\bibitem{wigner} Wigner, E.P., {\em Group Theory and its Applications to  
the Quantum Mechanics of Atomic Spectra} (Academic Press, N.Y., 1956).   
 
\bibitem{bargmann} Bargmann, V., {\em J. Math. Phys.} {\bf 5} (1964), 862--868. 
 
\bibitem{cassinelliB} Cassinelli, G., de Vito, E., Lahti, P.J. and Levrero, A., 
{\em Rev. Math. Phys.} {\bf 9} (1997), 921--941. 
 
\bibitem{cassinelli} Bracken, A.J., Cassinelli, G. and Wood, J.G.,   
{\em J. Phys. A}  {\bf 36} (2003), 1033-1056.   
  
\bibitem{tatarskii} Tatarskii, V., {\em Sov. Phys. Usp.} {\bf 26} (1983), 311--327. 
 
 
  
\end{thebibliography}
\end{document}